\documentclass[pdflatex,sn-mathphys-num]{sn-jnl}

\usepackage{graphicx}%
\usepackage{multirow}%
\usepackage{amsmath,amssymb,amsfonts}%
\usepackage{amsthm}%
\usepackage{mathrsfs}%
\usepackage[title]{appendix}%
\usepackage{xcolor}%
\usepackage{textcomp}%
\usepackage{manyfoot}%
\usepackage{booktabs}%
\usepackage{algorithm}%
\usepackage{algorithmicx}%
\usepackage{algpseudocode}%
\usepackage{listings}%
\usepackage{soul} 
\usepackage[mathlines]{lineno}
\usepackage{setspace}
\usepackage{geometry}

\usepackage{array} 

\theoremstyle{thmstyleone}%
\theoremstyle{thmstyletwo}%

\theoremstyle{thmstylethree}%

\begin{document}


\title[Article Title]{Quasi-monoenergetic Deuteron Acceleration via
Boosted Coulomb Explosion by Reflected Picosecond Laser Pulse}


\author[1]{\fnm{Tianyun} \sur{Wei}}\email{wei.tianyun@qst.go.jp}

\author[2]{\fnm{Zechen} \sur{Lan}}

\author[2]{\fnm{Yasunobu} \sur{Arikawa}}

\author[3]{\fnm{Yanjun} \sur{Gu}}

\author[1]{\fnm{Takehito} \sur{Hayakawa}}

\author[2]{\fnm{Alessio} \sur{Morace}}

\author[2]{\fnm{Ryuya} \sur{Yamada}}

\author[2]{\fnm{Kohei} \sur{Yamanoi}}

\author[2]{\fnm{Koichi} \sur{Honda}}

\author[1]{\fnm{Masaki} \sur{Kando}}

\author[1]{\fnm{Nobuhiko} \sur{Nakanii}}

\author[4]{\fnm{Seyed Reza} \sur{Mirfayzi}}

\author[5,1]{\fnm{Sergei Vladimirovich} \sur{Bulanov}}

\author[2]{\fnm{Akifumi} \sur{Yogo}} 

\affil[1]{\orgdiv{Kansai Institute for Photon Science (KPSI)}, \orgname{National Institutes for Quantum Science and Technology (QST)}, \orgaddress{ \city{Kizugawa}, \postcode{619-0215}, \state{Kyoto}, \country{Japan}}}

\affil[2]{\orgdiv{Institute of Laser Engineering}, \orgname{The University of Osaka}, \orgaddress{\city{Suita}, \postcode{567-0871}, \state{Osaka}, \country{Japan}}}

\affil[3]{\orgdiv{Institute of Scientific and Industrial Research (SANKEN)}, \orgname{The University of Osaka}, \orgaddress{\city{Ibaraki}, \postcode{567-0047}, \state{Osaka}, \country{Japan}}}

\affil[4]{\orgname{Tokamak Energy ltd}, \orgaddress{\city{Milton}, \postcode{OX14 4SD},  \country{United Kingdom}}}

\affil[5]{\orgname{Extreme Light Infrastructure ERIC ELI Beamlines Facility}, \orgaddress{\city{Dolní Břežany}, \postcode{252 41},\country{Czech Republic}}}


\abstract{

Generation of quasi-monoenergetic ions by intense laser is one of long-standing goals in laser-plasma physics. However, existing laser-driven ion acceleration schemes often produce broad energy spectra and limited control over ion species. Here we propose the acceleration mechanism, boosted Coulomb explosion, initiated by a standing wave, which is formed in a pre-expanded plasma by the interference between a continuously incoming main laser pulse and the pulse reflected by a solid target, where the pre-expanded plasma is formed from a thin layer on the solid target by a relatively strong pre-pulse. This mechanism produces a persistent Coulomb field on the target front side with field strengths on the order of TV/m for picoseconds. We experimentally demonstrate generation of quasi-monoenergetic deuterons up to 50 MeV using an \textit{in-situ} D$_2$O-deposited target. Our results show that the peak energy can be tuned by the laser pulse duration.

}




\maketitle

\large

The ability of producing ion sources with high brightness using ultra-intense lasers  \cite{daido2012review, macchi2013ion} has provided a platform to develop various applications
 such as compact neutron source \cite{kar2016beamed, yogo2023laser,lan2024single, mirfayzi2024recent}, fast ignitions \cite{roth2001fast,kodama2002fast, fernandez2009progress, fujioka2016fast}, proton radiography\cite{romagnani2008proton, shi2022picosecond, huang2025characterization} and radiobiological research\cite{yogo2011measurement, bin2012laser, bin2022new}. Protons with energies of 150~MeV have already been realized\cite{ziegler2024laser} which shows the possibilities of further applications such as cancer therapy \cite{bulanov2002feasibility,bulanov2014laser}. 
Many  ion acceleration mechanisms have been proposed, including Collisionless Shock Acceleration (CSA)\cite{silva2004proton, marques2024collisionless}, Radiation Pressure Acceleration (RPA)\cite{esirkepov2004highly, bin2015ion}, and Target Normal Sheath Acceleration (TNSA)\cite{wilks2001energetic,passoni2010target}. 
Among these mechanisms,  TNSA is widely applied in most laser systems. 
CSA and RPA are considered to be more efficient and reliable mechanisms rather than TNSA, but the requirement for the target thickness and laser intensity are limiting their application in current laser systems.

There are two remarkable features of TNSA mechanism. The first is that ions in a contamination layer on a target surface are  first accelerated mainly making it difficult to accelerate effectively other type of ions rather than protons.
Thus, previous studies have proposed some approaches to accelerate other type of ions in TNSA  such as using a cryogenic D$_2$ target to realize pure deuteron acceleration\cite{wei2024cryogenic}, heating target surface by additional laser for heavy ion acceleration\cite{wang2021super}, and focusing an ultra intense laser on a nanometer scale target\cite{nishiuchi2020dynamics}.
The second  is that the accelerated ions have typically  a continuous distribution,  which are not suitable for some applications such as the study of nuclear physics and nuclear engineering. 
For example, monoenergetic deuterons via high power laser can be used for neutron sources with selective energy using the secondary reactions such as the $d$+$^9$Be reaction,\cite{Wei13}
,the study of $d$+$d$ and $d$+$t$ nuclear fusion \cite{gatu2017development, schwemmlein2022first} , and production of medical radioisotopes \cite{Daraban07}

Producing quasi-monoenergetic particles using lasers \cite{winkler2025active} remains a significant challenge. 
Many efforts \cite{palaniyappan2015efficient, jung2011monoenergetic, henig2009radiation, esirkepov2002proposed, schwoerer2006laser, morita2008tunable, ahmed2021high, ter2006quasimonoenergetic, ramakrishna2010laser, bulanov2008accelerating, morita2014approach, wei2024realizing} including different acceleration mechanisms have been made to generate quasi-monoenergetic ions, for example, using the ultra-thin target\cite{palaniyappan2015efficient, jung2011monoenergetic, henig2009radiation}, the microstructured target \cite{bulanov2002feasibility, esirkepov2002proposed, schwoerer2006laser, morita2008tunable} and the  coil  target\cite{ahmed2021high}.
The simplest method is the use the ultra-thin target.
When the thickness of the thin layer is thin enough to be collectively affected by the laser filed, quasi-monoenergetic
ions could be accelerated. 

In our recent study \cite{wei2024realizing},  we demonstrated a method for the fabrication of \textit{in-situ} D$_2$O deposited targets, achieving accleration of quasi-monoenergetic deuterons ($\Delta E/E = 4.6\%$) through TNSA mechanism. However, protons play significant roles in the formation of quasi-monoenergy component of deuterons. The electric field generated by the accelerated protons suppress the energies of the high-energy deuterons to form quasi-monoenergy deuterons. This suppression limits the peak energy of the deuterons to approximately 11 MeV. When we use a target with a proton layer, it is difficult to accelerate deuterons much higher than 11 MeV. Thus, we need alternative mechanism to accelerate quasi-monoenergetic high-energy deuterons.

Coulomb explosion\cite{ter2006quasimonoenergetic, bulanov2008accelerating, ramakrishna2010laser, morita2014approach} is another possible mechanism to generate quasi-monoenergetic ions, which has the potentiality to generate ions with peak energy of several tens of MeV in the backward direction of the laser \cite{gu2012large}. 
A combination of the charge separation electric field and the field of the laser can result in generation of quasi-monoenergetic proton beams in so-called the "directed Coulomb explosion acceleration regime"\cite{bulanov2008accelerating}.
However, because of the low efficiency of Coulomb explosion, the energies of accelerated ions are limited to only 1$\sim$2~MeV in the previous experiments \cite{ter2006quasimonoenergetic, ramakrishna2010laser}. 
In addition, it has been theoretically proposed that protons could be accelerated up to 100~MeV through a longitudinal charge-separation field generated by chirped standing waves formed by a fs laser pulse reflected by a high-density mirror located behind a target \cite{mackenroth2016chirped} .

In this study, we investigate that a standing wave generated by reflecting a picosecond laser pulse with the critical surface of a plasma \cite{smith2019particle} could enhance the efficiency of the Coulomb explosion for producing a persistent electric field.
We conduct ion acceleration experiment using Al targets with ultra-thin D$_2$O layer targets using the \textit{in-situ} fabricated method. We introduce a pre-pulse to expand the ultra-thin D$_2$O layer before the laser main pulse. As results, we measured high energy quasi-monoenergetic deuterons with energies of up to 50~MeV.
Simulation results reveal the mechanism that dominates the acceleration of the quasi-monochromatic deuterons. A relatively strong pre-pulse expands a thin D$_2$O layer on a solid target to form a underdense plasma, and subsequently a picosecond main laser pulse can penetrate through this plasma but is reflected by the solid density region.
The reflected pulse and the continuously incident laser pulse then form a standing wave at the front of the target, which effectively kicks out electrons and generates a strong Coulomb explosion field. This field, on the order of TV/m, persists for several picoseconds and enables the acceleration of deuterons to quasi-monoenergetic spectra with peak energies reaching tens of MeV.

\section*{Results}

\begin{figure*}[th]
\centering
\includegraphics[width=16cm]{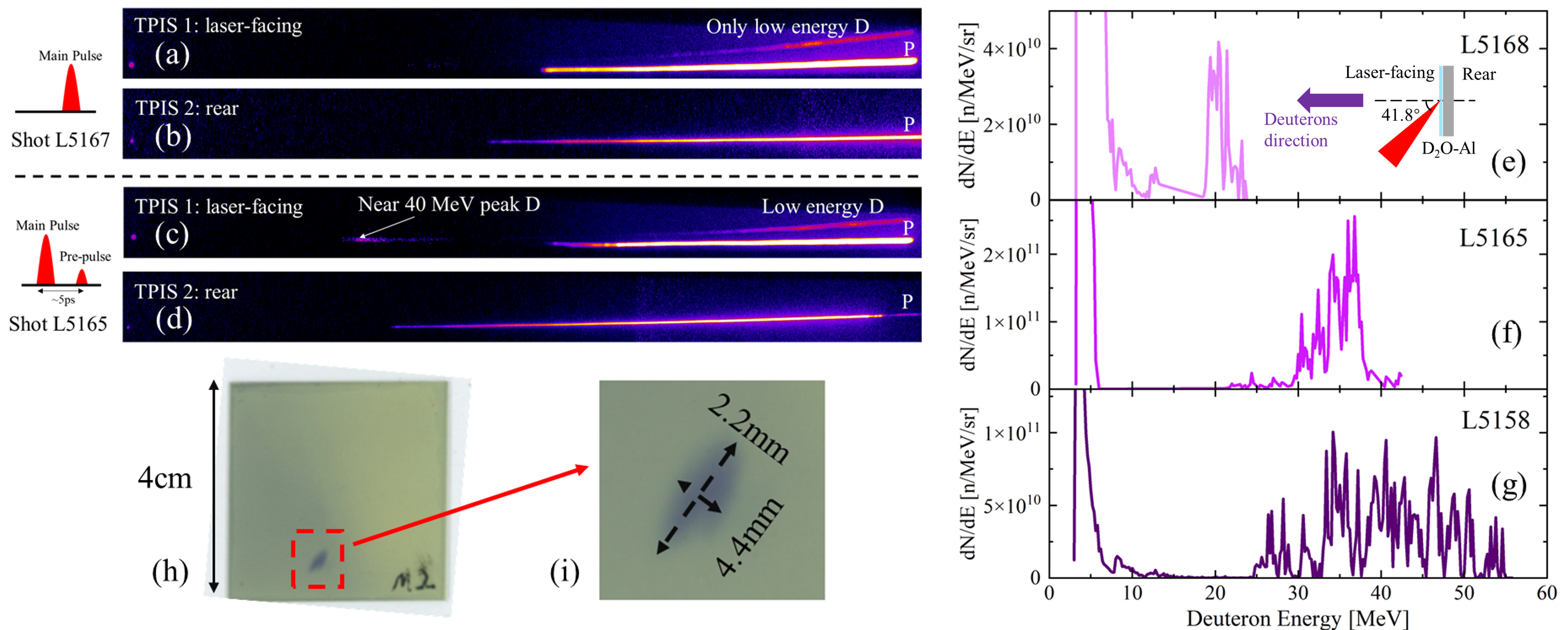}
\caption{
(a)-(b) Raw data of TPISs for shot L5167. Only low-energy deuterons are measured on the target laser-facing side.
(c)-(d) Raw data of TPISs for shot L5165. Shot with pre-pulse, $\sim$ 40~MeV quasi-monoenergetic deuterons are measured at the laser-facing side.
(e)-(g) The deuteron spectra at laser-facing side for 3 different shots. Structured deuterons from 20 to 50~MeV are measured.
(h) The measured highest energy profile by RCF, the corresponding deuteron
energy is 33.0~MeV.
(i) The spot is nearly an ellipse with size of 4.4~mm$\times$2.2~mm.
}
\label{fig_tpis_reulst}
\end{figure*}

The typical raw data of the Thomson Parabola Ion Spectrometers (TPISs) \cite{tosaki2017evaluation} for the shots with only main pulse are shown in Fig.\ref{fig_tpis_reulst}
(a) and (b). There are  low energy deuterons measured only at the  laser-facing side of the target. In contrast, the typical raw data of  the TPISs for the shots with pre-pulse are shown in Fig.\ref{fig_tpis_reulst}(c)  and (d). 
Deuterons are still only measured at the laser-facing side, where the peak energy of the deuterons is approximately 40~MeV. The energy spectras of 3 different shots with pre-pulse at laser-facing side are shown in Fig.\ref{fig_tpis_reulst}(e)-(g).
Each energy spectrum is obtained from a single laser shot without accumulation.
 The peak energies vary from 20~MeV to 50~MeV. This result indicates that we may control the deuteron peak energy by optimizing the thickness of D$_2$O layer and the laser parameters. Note that the TPISs used in the experiments have an energy resolution better than 0.2~MeV in the 3-30~MeV range. At higher energies ($>$30 MeV), the resolution decreases, which may broaden the observed spectral features.

\begin{figure*}[th]
\centering
\includegraphics[width=8cm]{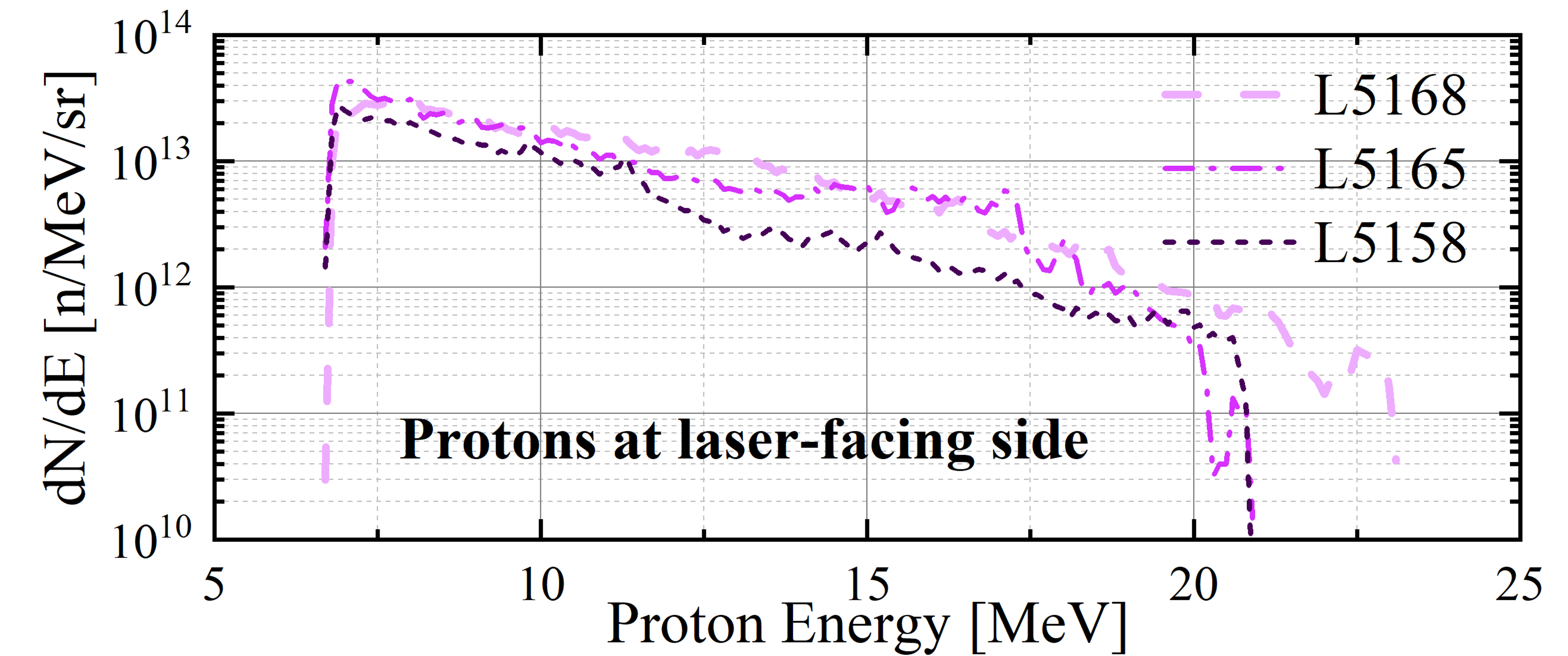}
\caption{
The proton spectra at laser-facing side for the same 3 shots with TPIS
}
\label{p_spec}
\end{figure*}

The spatial  pattern of the highest energy deuterons measured by the Radiochromic Films (RCF)\cite{nurnberg2009radiochromic} in the beam profile shot is shown in Fig.\ref{fig_tpis_reulst}(h).
By calculating the particle transport with the PHITS Monte Carlo simulation code\cite{sato2024recent}, the minimum energies of the protons and deuterons reaching to the RCF layer presented in Figs.~\ref{fig_tpis_reulst}(h) and (i) through other layers is 24.6~MeV and 33.0~MeV, respectively.  Because the maximum energy of the protons is 23 MeV as shown in Fig. \ref{p_spec}, the signals in the RCF originate only from to the deuterons.
It is difficult to evaluate the deuteron energy spectrum lower than 33.0~MeV from the RCF stacks, because both protons and deuterons make signals in the lower energy RCF layers.
Because the laser irradiates on the target with an angle from the normal direction of the surface, the deuteron profile shows a nearly ellipse spot with a size of 4.4~mm$\times$2.2~mm. 
By considering the distance of the RCF from the target (31 mm), the divergence of the high-energy deuterons is estimated as $\Omega=\pi\times3.3\times4.4/31^2\approx0.03$~sr. Assuming an axially symmetric conical distribution, the half angle $\theta\approx5.6^{\circ}$ is derived from $\Omega=2\pi(1-cos\theta)$.

In the previous work\cite{wei2024realizing}, we accelerated quasi-monoenergic deuterons near 11~MeV using a picosecond laser pulse with an energy of approximately 600~J. From an ultra thin D$_2$O layer (tens of nanometers) at the rear side of the target (no D$_2$O layer in the laser-facing side), deuterons are accelerated early by the TNSA mechanism, and protons are also accelerated later. High and low velocity protons compress the deuteron distribution from both velocity regions in the momentum space, and thereby a component of quasi-monoenergic deuterons is formed. In contrast, in the present study, the deuterons are accelerated from the laser-facing side of the target and their energies are higher than those of the simultaneously accelerated protons. These facts indicate that the quasi-monoenergic deuterons are accelerated by a mechanism different from the TNSA in the previous study. 
Note that the detection threshold of the TPIS is approximately 3~MeV for deuterons. 

\begin{figure*}[th]
\centering
\includegraphics[width=16cm]{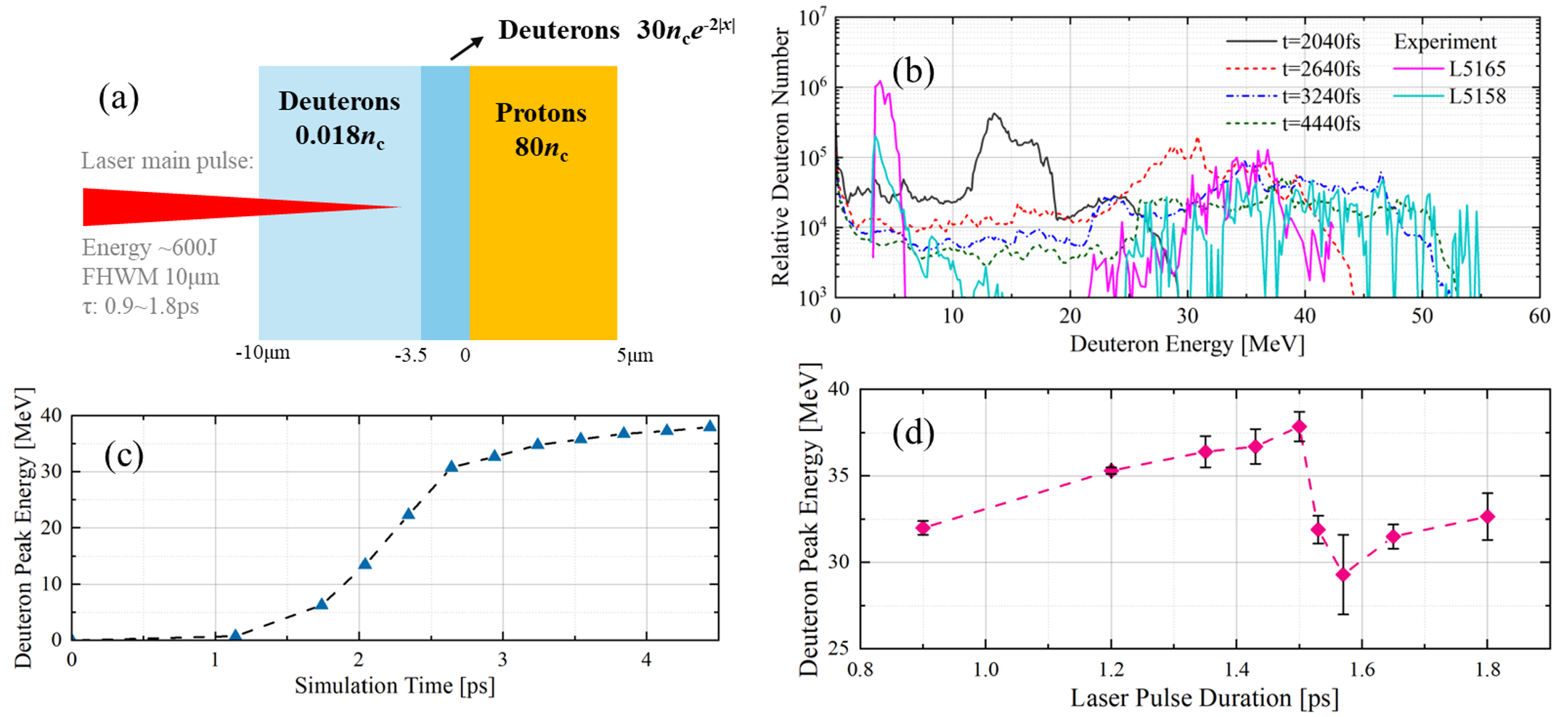}
\caption{
(a)Simulation set-up, self-focused main pulse incident into the solid target with pre-expanded deuterons. (b) Energy spectra at different simulation time. Experimental spectra of L5165 and L5158 are also included.(c) The deuteron peak energy time evolution. (d) Laser pulse duration dependence of the deuteron peak energy.
}
\label{sim}
\end{figure*}

To understand the acceleration process of the structured deuterons at the laser-facing side of the target, we conduct 2D Paricle-In-Cell (PIC) simulations with the EPOCH code\cite{bennett2017users} using the set-up shown in Fig.\ref{sim}(a). 
The energy spectra of the deuterons at the laser-facing side of the target in a time range from t=2.04~ps to t=4.44~ps are shown in Fig.\ref{sim}(b). 
The energy spectra measured in the shot L5165 and L5158 are also included in Fig. 2(b) to illustrate the agreement between simulation and experiment results.
The deuterons shows a 
structured component around 15~MeV which appears at 2.04~ps. The energy of this component is further increased and become stable at  3~ps. Fig. \ref{sim}(c) shows the evolution of the deuteron peak , where its energy starts to increase after 1~ps, and  rapidly increases up to 2.7~ps. It finally become stable after 3~ps with a peak energy  of approximately 35~MeV. 

Fig.~\ref{sim}(d) shows the deuteron peak energy dependence on the pulse duration of  the incident laser. The pulse durations are changed from 0.8~ps to 1.8~ps while  the incident laser energies remain as a constant by  decreasing the laser intensity. The bandwidth of the peak components are adopted as the error bar for the peak energy. 
The calculated result [Fig.\ref{sim}(d)] shows that the maximum peak energy of the accelerated deuterons is obtained in the case of a pulse duration of 1.5~ps, which is nearly equal to the typical pulse width of the LFEX laser. If we use laser pulses longer than 1.5~ps, the peak intensity and normalized vector potential $a_0$ decrease ($a_0\propto1/\sqrt{\tau}$) under the condition that the total laser pulse energy is fixed. As $a_0$ decreases, the penetration depth of a laser pulse in a relativistic plasma (relativistic skin depth) shortens and the amplitude of the reflected laser to form a standing wave decreases. These effects reduce the strength of the boosted Coulomb explosion field to accelerate deuterons. As shown in Fig.~\ref{sim}(d), when pulse widths are longer than 1.5~ps, the peak energies of the accelerated deuterons are lower than that at 1.5~ps by approximately 5--8~MeV. The energy spectrum of the accelerated deuterons depends on the density and the scale length of the pre-plasma. In the present condition, the pre-expanded D$_2$O layer has an exponential profile with a few-micrometer scale length. If the pre-plasma has a thin scale length or low density, a long laser pulse penetrates easily the pre-plasma and accelerates effectively deuterons. Conversely, when the pre-plasma has a thick scale length or a high density, a shorter laser pulse is suitable for effective deuteron acceleration.

\begin{figure*}[th]
\centering
\includegraphics[width=16cm]{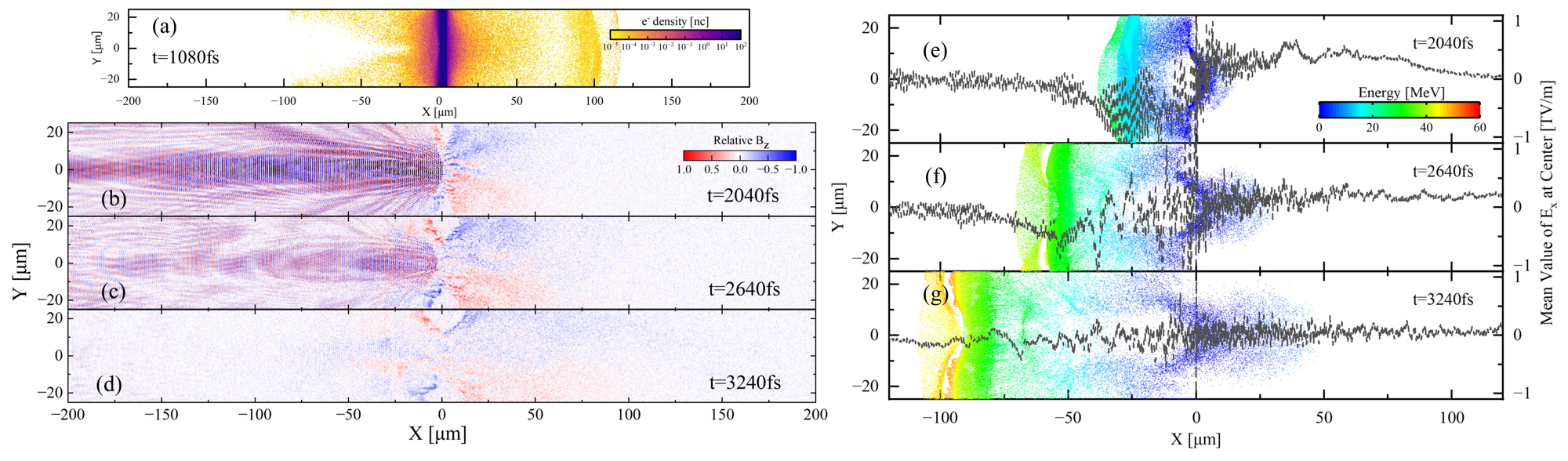}
\caption{
(a)Electron density while laser incident on the solid target. Electrons are kicked off by the laser pulse at the laser-facing side of target. (b)-(d) The B$_z$ field, indicate the laser pulse position during simulation. (e)-(g)The deuteron energy distribution and E$_x$ field at the center, deuterons are accelerated by a TV/m level Coulomb Explosion field lasting multi-picoseconds. 
}
\label{sim_2D}
\end{figure*}

The electron density while the laser main pulse incidents on the solid target (1.08ps) is shown in Fig.\ref{sim_2D}(a). The electrons are heated and transparented into the target rear side, forming the sheath field at the target rear. In contrast, at the laser-facing side, electrons are quickly kicked out by the laser pulse. As a result, a strong Coulomb explosion field is formed at the target front side by the left positive charged ions. The Figs.~\ref{sim_2D}(b)-(d) show the 2D distribution of B$_z$ indicating the position of laser pulse. The time stamps in these figures show the simulation time. The laser pulse is set from the left boundary (-200~$\mu$m) with a pulse width of 1.5~ps and a peak at 1~ps. This pulse reaches the surface of the target at approximately 0.67~ps later, and the peak of the pulse reaches this surface at approximately 1.67~ps.

Figs.\ref{sim_2D}(e)-(g) show the 2D energy distribution of the deuterons, and the black dashed lines show the spatial mean value of E$_x$ field at the center (-2.5~$\mu$m to 2.5~$\mu$m) .
At early time (2.04~ps) the laser incidents on the solid target, and the laser pulse is reflected by the solid target. The reflected pulse and the continuous incident pulse  form a standing wave that kicks out the electrons effectively to lead the Coulomb explosion field at the laser-facing side. The deuterons at the laser-facing side are rapidly accelerated by this field. At a later time ( 2.64~ps), the incidence of the laser is already finished, but the reflected laser pulse  kicks out the electrons  and maintains the Coulomb explosion field. This field exists just behind the highest energy deuterons, and accelerates  deuterons during a sub-picosecond duration.  Once the laser including the reflected laser has completely disappeared at  3.24~ps, the electrons expand to the laser facing side, and the process is terminated, and deuteron energy is becoming stable. 
The Coulomb explosion field maintains its strength in the order of $\sim$~TV/m for multi-picoseconds after which the laser incidents on the solid target because the  main and reflected laser pulses  kick out the electrons in multi-picoseconds. 
The electric field generated by electrons at the laser-facing side could be estimated from the electron density distribution shown in Fig.~\ref{sim_2D}(a).
The estimated field is on the order of 10$^9$~V/m, which is much lower than the Coulomb explosion field strength in order of TV/m.
This Coulomb explosion field accelerates the deuterons to the maximum energy of $\sim$50~MeV. The acceleration field exits behind the highest energy deuterons, and there are no electrons that form other electric field to expand the deuteron energy so that the deuterons  keep the peak energy during the acceleration phase. In addition, the accelerated deuterons have an arc shape, which is formed  by the hole boring effect \cite{kodama1996study} of the laser main pulse. This allows the highest energy deuterons to collimate themselves around the center along the target normal direction.  This is consistent with measured small divergence. 

Note that multi-peak structures are observed in all three experimental shots [Fig.\ref{fig_tpis_reulst}(e)-(g)]. During the early stage when the standing wave forms at tens of femtoseconds, electron modulation may induce a multi-peak structure on the deuterons \cite{mackenroth2016chirped} before the main Coulomb explosion acceleration.

\begin{figure*}[th]
\centering
\includegraphics[width=16cm]{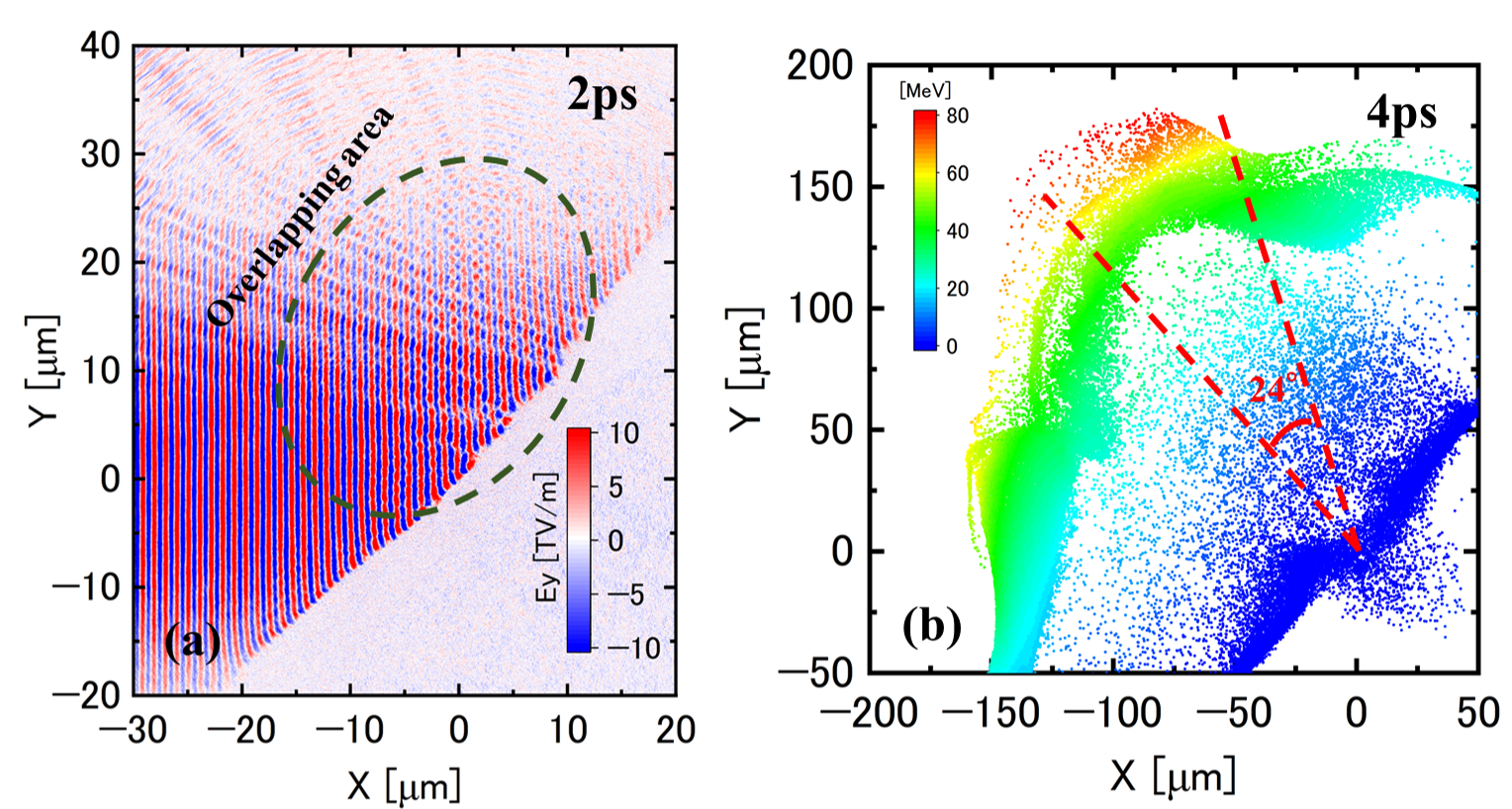}
\caption{
(a)Simulations for oblique incidence, standing wave is formed at the overlapping area of incident and reflected pulses. (b)Deuterons are able to accelerated into several tens of MeV on the laser-facing side at normal direction.
}
\label{sim_angle}
\end{figure*}

In the above simulations, the angle of the incident laser has been assumed to be 0$^{\circ}$ from the normal direction of the target surface. Because the incident angle in the experiment is 41.8$^\circ$, we perform an additional simulation where the incident angle of 42$^\circ$ is assumed. Figure~\ref{sim_angle}(a) presents that a standing wave still forms in the overlap region of the incident and reflected pulses and the quasi-monoenergetic deuterons are accelerated up to approximately 80~MeV along the normal direction with a open divergence angle of smaller than 24$^\circ$ shown in Fig.~\ref{sim_angle}(b). The standing wave leads to localized electron evacuation, and thus a strong electrostatic field is generated along the normal direction. This accelerates deuterons preferentially along the normal direction.

The simulation result shows clearly that the boosted Coulomb explosion mechanism dominates the acceleration of the  deuterons in the quasi-monoenergetic peak at the laser-facing side. There are two key points for this mechanism. The first is that a thin pre-plsama should exsit on the laser-facing side of a target, allows the incident laser has a large skin-depth to kick the electrons out of the target to create a strong Coulomb explosion field in the order of TV/m. 
In our experiments, to make such thin pre-plasma, we deposited target in-situ. This method creates a ultra-thin D$_2$O layer which could be pre-expanded to a thin plasma by a pre-pulse with a fraction of $\sim$5$\%$ of the main pulse.
The second  is that a relatively longer laser pulse  to maintain the field in a multi-picoseconds.  The LFEX laser has the  pulse duration of   $\sim$1.5~ps, which allows the acceleration field could last to $\sim$ 2~ps because the reflect laser could also kick out the electrons to maintain the Coulomb Explosion field. 
This method could apply to other picosecond lasers such as OMEGA EP \cite{kelly2006omega} and PETAL \cite{casner2015lmj} for quasi-monoenergetic ion acceleration under energy control.

Finally, we discuss a possibility that we form a boosted Coulomb explosion to other ion species such as tritons and protons. The accelerated ions could be selected by depositing a contamination layer on a metal target. 
We use D$_2$O deposited target for deuteron acceleration, it is acceptable to realize quasi-monoenegetic triton acceleration which is expected in research such as nuclear fusion avoding the possible high cost of isotopes and radiation hazards in traditional accelerators \cite{schwemmlein2022first} .
As for quasi-monoenergetic proton acceleration which is necessary for applications such as cancer therapy, because it is difficult to make a ultra-thin layer of protons on a metal target, we suggest the use of a stronger pre-pulse to expand quickly the proton plasma.


In this work, we investigated the boosted Coulomb explosion mechanism as a means of accelerating ions to energies of several tens of MeV using picosecond laser pulses. Through this mechanism, we experimentally achieved quasi-monoenergetic deuterons with peak energies up to 50 MeV by irradiating \textit{in-situ} D$_2$O-deposited targets with the LFEX laser system.
A thin D$_2$O layer was formed on an aluminum target by the evaporation of D$_2$O molecules through nanoscale holes in a plastic capsule containing liquid D$_2$O. This layer was pre-expanded by a pre-pulse, enabling the main laser pulse to transmit through the underdense plasma and reflect by the solid aluminum surface.
Ion acceleration at the laser-facing side is initiated by a standing wave formed by the interference between the reflected and continuously incident picosecond laser pulses. This standing wave efficiently expels electrons, generating a strong Coulomb explosion field on the order of TV/m, persisting for several picoseconds.
Simulations show that ions can be accelerated to tens of MeV with a narrow energy spread, and that the peak energy can be tuned by adjusting the laser pulse duration. 
The boosted Coulomb explosion mechanism provides an energy controllable approach for generating quasi-monoenergetic ions using high power laser systems. Moreover, the ion species can potentially be selected by depositing appropriate surface layers on the target, offering flexibility in tailoring ion sources. This mechanism holds strong promise for a wide range of applications, including fast ignition in inertial confinement fusion, nuclear physics experiments, and laser-driven cancer therapy.

\section*{Methods}

\subsection*{Experiments}

The experiments are conducted with the Laser for Fast Ignition Experiment (LFEX) system \cite{kawanaka20083} at the Institute of Laser Engineering in the University of Osaka. 
The LFEX laser provides three shots per day and four pulses of H1--H4 with a center wavelength of 1.05~$\mu$m are simultaneously delivered for each shot. The focal spot diameters of the four pulses are 50~$\mu$m for H1, H3, and H4 and 30~$\mu$m for H2. The laser pulse energy and duration of each shot used in the experiments are listed in Table~\ref{laser_para}.
The laser is operated in two modes. The first mode is normal shot with main pulse only, and the second mode is shot with a pre-pulse approximately 5~ps before the main pulse. The energy of the pre-pulse is approximately 5$\%$ of the main pulse. 
For all the shots with pre-pulse, we have high-energy components of deuterons at laser-facing side which indicate a high reproducibility.

\begin{figure*}[th]
\centering
\includegraphics[width=16cm]{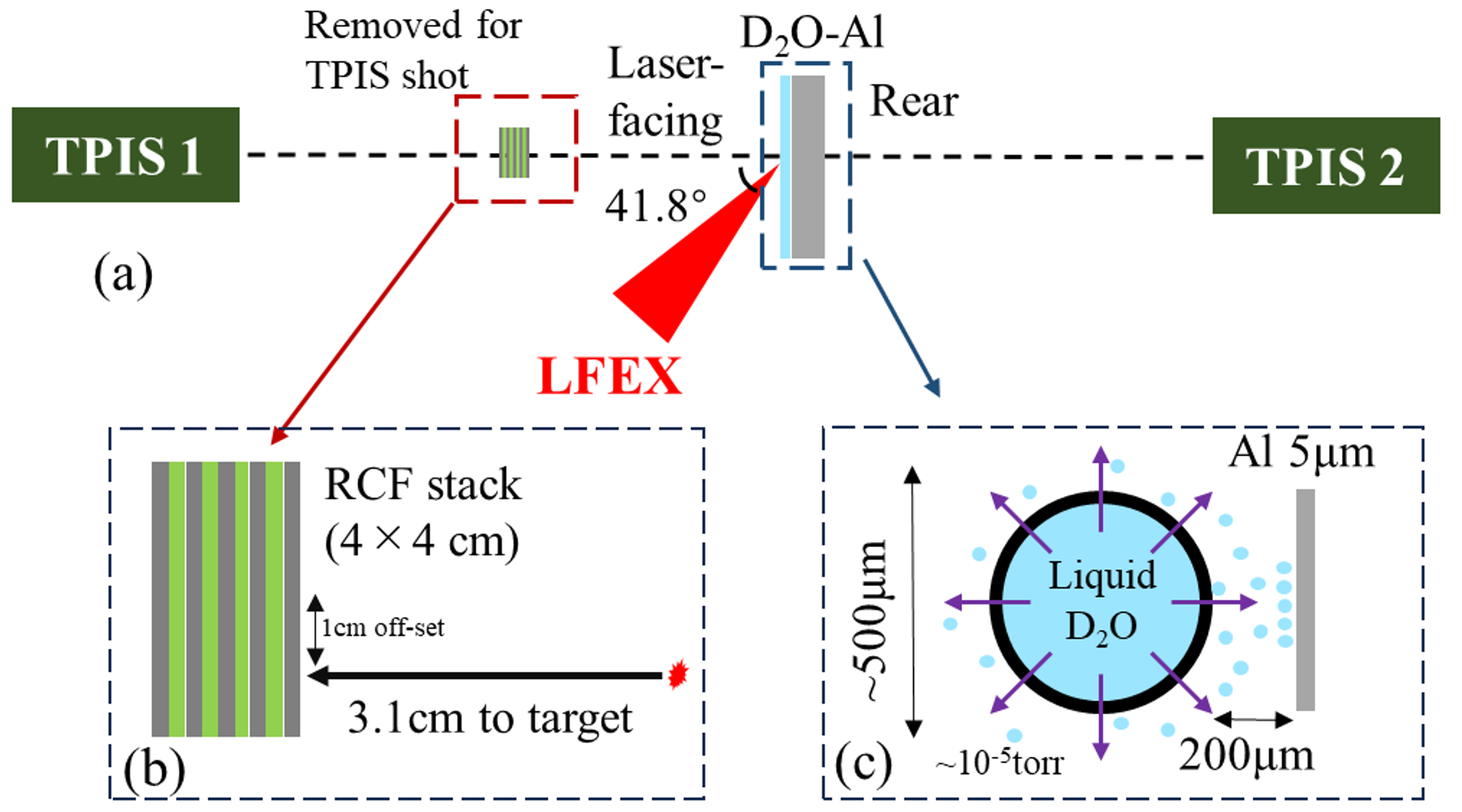}
\caption{
(a) Deuteron acceleration experiment set-up with the D$_2$O-Al target using the LFEX laser. 
(b) RCF stack set 3.1~cm at front of the target in deuteron beam profile shot.
(c) D$_2$O-Al target \textit{in-situ} fabricated using D$_2$O capsule in vacuum chamber ($\sim$10$^{-5}$~torr).
}
\label{fig_tpis_set}
\end{figure*}

\begin{table}[]
\centering
\caption{The shot-by-shot laser energy and pulse duration}
\begin{tabular}{cccccccccccc}
\hline
\multirow{2}{*}{Shot No.} & \multirow{2}{*}{Pre-pulse} & \multicolumn{4}{c}{Pulse FWHM (ps)} & \multicolumn{5}{c}{Laser energy on target (J)} & \multirow{2}{*}{\begin{tabular}[c]{@{}c@{}}D$^{+}$ peak\\ (MeV)\end{tabular}} \\ 
                          &                            & H1      & H2      & H3     & H4     & H1      & H2      & H3     & H4     & Total    &                                                                         \\ \hline
L5158                     & w                          & 1.89    & 1.80    & 1.53   & 1.73   & 160     & 163     & 162    & 118    & 603      & 40.6                                                                    \\
L5165                     & w                       & 1.92    & 1.78    & 1.83   & 1.68   & 182     & 190     & 189    & 138    & 699      & 34.6                                                                       \\
L5167                     & w/o                          & 1.87    & 1.74    & 1.44   & 1.60   & 167     & 181     & 167    & 133    & 648      & -                                                                    \\
L5168                     & w                          & 1.95    & 1.82    & 1.68   & 1.66   & 173     & 175     & 175    & 128    & 651      & 20.4                                                                    \\
L5169                     & w                          & 1.72    & 1.60    & 1.53   & 1.44   & 192     & 194     & 188    & 143    & 717      & 33.0                                                                    \\
L5171                     & w/o                        & 1.74       & 1.61       & 1.59      & 1.44      & 182       & 184       & 181      & 134      & 681        & -                                                                       \\ \hline
\end{tabular}
\label{laser_para}
\end{table}

The experiment set-up is shown in Fig.\ref{fig_tpis_set}(a). The LFEX laser irradiates on the D$_2$O-Al target front side with 41.8$^{\circ}$, two TPISs are set on the front side and rear side to identify ion species and measure the energy spectra of the accelerated ions. The Al filters with a thickness of max 300 $\mu$m are used in the TPISs to stop heavy ions such as carbons and oxygen so that only protons and deuterons are measured with the TIPSs. The RCF and Al filter stack with the size of 4$\times$4~cm is set at the front side of the target with a distance of 3.1~cm in the deuteron pulse profile shot after the measurement of the energy spectra with TPISs. 
The RCF stack was placed parallel to the target surface but offset by approximately 1~cm from the centerline to avoid interference with the laser beam.
Each Al filter is inserted between two RCFs to decrease the deuteron energy.
A D$_2$O-Al target \cite{wei2024realizing} is \textit{in-situ} fabricated using a D$_2$O capsule to deposit a D$_2$O layer on the surface of an Al target  as shown in Fig.\ref{fig_tpis_set}(c).
Before a laser shot, we locate a spherical plastic capsule including D$_2$O water near an Al target. A part of D$_2$O molecules evaporate through nano-scale holes of the plastic capsule, and subsequently accumulate on the surface of the Al target so that a thin D$_2$O layer is formed. The thickness of the D$_2$O layer is approximately proportional to the depositing time. 
The depositing times of $\sim$40~min are constant for every shot to make every target having almost the same thickness of the D$_2$O layer. The D$_2$O capsules are removed from the target chamber before laser shots. The Al targets are kept at near room temperature (25$^\circ$C) during the deposition process.

\subsection*{Simulations}
The simulation box is in the region of from -200 to 200~$\mu$m for the x-axis and from -25 to 25~$\mu$m for the y-axis with a mesh size of 50~nm. 
As shown in Fig.~\ref{sim}(a) protons with a density of 80~$n_c$ are set in the solid Al target region from 0 to 5~$\mu$m for the x-axis. The protons are used instead of Al ions because of computational efficiency. By considering a laser pre-pulse, low-density deuteron plasma with a density of 0.018~$n_c$ is set in the region from -10 to -3.5~$\mu$m for the x-axis to present the pre-pulse expansion, which is estimated to have a temperature of 10 keV over 5~ps.
High-density deuterons is set in the region from -3.5 to 0~$\mu$m for the x-axis to represent
 the fully ionized D$_2$O layer ($\sim33~n_c$).
This deuteron layer has an exponential distribution of 30~$n_ce^{-2\lvert x\rvert}$ at the laser-facing side of the proton layer.
Note that oxygen ions are not set in the deuteron plasma because of computational efficiency.
A laser pulse with a total energy of 600~J (typical total energy of LFEX) incidents into the target from the thin deuteron plasma side. The laser pulse duration is 1.5~ps in FWHM in the simulations. The wavelength is 1.05~$\mu$m identical with the experimental parameter.
The laser spot size is set to 10~$\mu$m in FWHM instead of the size in the experiments of 50~$\mu$m to account for self-focusing effects and to reduce computational complexity. 
The laser intensity of the simulation increases from the experiment intensity of 10$^{19}$W/cm$^2$ to approximately 2.5$\times$10$^{20}$W/cm$^2$. 
This higher intensity allows deeper penetration of the main pulse into the pre-plasma, but the  reflected surface is shifted to a position of a higher electron density of $n_c\sqrt{1+a^2_0/2}$ which is approximately 4.5 times higher than that in the experiments.
Since the focal spot size is 5 times smaller in the horizontal direction, the total number of evacuated electrons remains comparable to the experimental case. As a result, the total number of evacuated electrons in the simulation is comparable to that in the experiments. This leads that the present simulation giving the Coulomb explosion field strength similar to the experimental condition.

The simulation box for the oblique incidence in Fig.~
\ref{sim_angle} is in the region of from -200 to 50~$\mu$m for the x-axis and from -100 to 300~$\mu$m for the y-axis with a mesh size of 50~nm. The laser spot (FWHM) is set as 25~$\mu$m with the same laser energy. The target set-up is the same with that for the normal incidence.

\section*{Acknowledgements}

The authors thank the technical support staff of ILE for their assistance with the laser operation, target fabrication and plasma diagnostics. This work was supported by the Collaboration Research Program of ILE, The University of Osaka(2023A1-004WEI). 
This work was funded by JSPS KAKENHI Grant-in-Aid for Scientific Research (JP25420911, JP26246043, JP22H02007, JP22H01239, 25K23369), JST A-STEP (AS2721002c) and JST PRESTO (JPMJPR15PD).
T. Wei was supported by JST SPRING (JPMJSP2138) during PhD before Sep. 2024.
T. Wei is now supported by KAKENHI International Leading Research FiMEC (23K20038).
The EPOCH simulations were conducted with the supercomputer HPE SGI8600 in the National Institutes for Quantum Science and Technology (QST). 
YJ. Gu acknowledges the use of the supercomputer facility of the Cybermedia Center at the University of Osaka.
Portions of this research were carried out (by S. V. B.) at the ELI Beamlines Facility  a European user facility operated by the Extreme Light Infrastructure ERIC.
The authors specially indebted to Prof. K. Mima who passed away for the initially discussions on the theories. 

\section*{Author contributions}
T. W., Z. L., Y. A., T. H., A. M. and R. Y. performed the experiments.
T. W. and  Z. L. analyzed the experimental data.
T. W. conducted the simulations.
Y. G. discussed on the theories and gave the advices on the simulations.
K. Y. and K. H. fabricated the capsule targets.
M. K. and N. N. discussed on the theories.
S. R. M. gave the original ideas of using heavy water and reviewed  the paper.
S. V. B. reviewed  the paper from theoretical viewpoints.
The manuscript was prepared by T. W. and T. H..
All authors contributed to discussions and the preparation of the manuscript.
A. Y. provided overall supervision of the work. 

\section*{Competing interests}
The authors declare no competing interests.

\section*{Data availability}
The data that support the findings of this study are available on request from the corresponding authors.

\bibliography{sn-bibliography}

\end{document}